\documentclass{elsarticle}
\usepackage{microtype}
\usepackage{graphicx}
\usepackage{subfigure}
\usepackage{booktabs}
\usepackage{multirow}
\usepackage{xcolor}
\usepackage{hyperref}
\usepackage{mathtools,cases}
\usepackage{amsfonts}
\usepackage{float}
\allowdisplaybreaks

\hypersetup{pdfauthor={Name}}

\def\nmo{N$-$1}%

\begin{document}

\title{Learning an Optimally Reduced Formulation of OPF through Meta-optimization}

\author[1]{Alex~Robson}
\author[1]{Mahdi~Jamei}
\author[1]{Cozmin~Ududec}
\author[1]{Letif~Mones\corref{cor}}

\address[1]{Invenia Labs, 95 Regent Street, Cambridge, CB2 1AW, United Kingdom (e-mails: \{firstname.lastname\}@invenialabs.co.uk).}
\cortext[cor]{Corresponding author}
\fntext[fn]{The authors thank Dr. Christian Steinruecken and Dr. Lyndon White for their suggestions that greatly improved the manuscript.}

\begin{abstract}
With increasing share of renewables in power generation mix, system operators would need to run Optimal Power Flow (OPF) problems closer to real-time to better manage uncertainty.
Given that OPF is an expensive optimization problem to solve, shifting computational effort away from real-time to offline training by machine learning techniques has become an intense research area.
In this paper, we introduce a method for solving OPF problems, which can substantially reduce solve times of the two-step hybrid techniques that comprise of a neural network with a subsequent OPF step guaranteeing optimal solutions.
A neural network that predicts the binding status of constraints of the system is used to generate an initial reduced OPF problem, defined by removing the predicted non-binding constraints.
This reduced model is then extended in an iterative manner until guaranteeing an optimal solution to the full OPF problem.
The classifier is trained using a meta-loss objective, defined by the total computational cost of solving the reduced OPF problems constructed during the iterative procedure.
Using a wide range of DC- and AC-OPF problems, we demonstrate that optimizing this meta-loss objective results in a classifier that significantly outperforms conventional loss functions used to train neural network classifiers.
We also provide an extensive analysis of the investigated grids as well as an empirical limit of  performance of machine learning techniques providing optimal OPF solutions.
\end{abstract}


\maketitle

\section{Introduction}
\label{sec:intro}
A central task of electricity grid operators \cite{Tong04} is to frequently solve some form of Optimal Power Flow (OPF) \cite{cain2012history}, which is a constrained optimization problem. 
The goal of OPF is to dispatch generation in order to meet demand at minimal cost, while respecting reliability and security constraints. 
This is a challenging problem for several reasons. 
First, OPF is a non-convex and non-linear constrained optimization problem that can take a mixed-integer form when solving the unit commitment problem. 
Second, it is computationally expensive due to the size of power grids, requiring a large number of diverse constraints to be satisfied. 
Further, grid operators must typically meet (at least) \nmo{} reliability requirements (e.g.~North American Electric Reliability Cooperation requirement for the US grid operators \cite{datanerc}), resulting in additional constraints that significantly increase the computational complexity of OPF.
Finally, with increasing uncertainty in grid conditions due to the integration of renewable resources (such as wind and solar), OPF problems need to be solved near real-time to have the most accurate inputs reflecting the latest state of the system. 
This, in turn, requires the grid operators to have the computational capacity of running consecutive instances of OPF problems with fast convergence time.

\begin{figure}[H]
    \centerline{\includegraphics[width=1.0\textwidth]{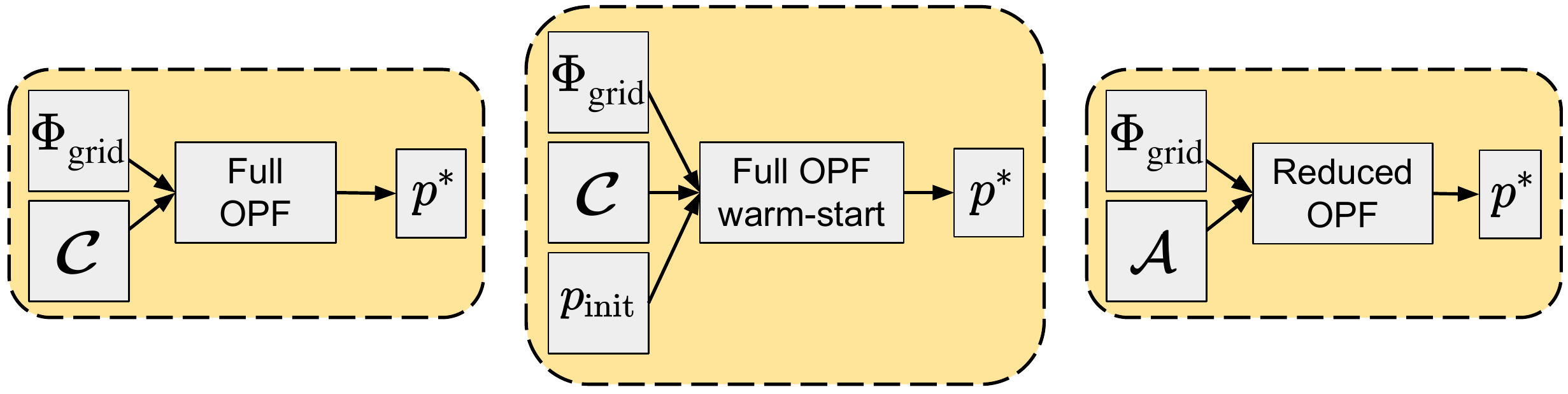}}
    \caption{OPF solution strategies for interior-point methods: conventional (left), warm-start (middle), and reduced (right) techniques. $\Phi_{\mathrm{grid}}$ denotes the vector of grid parameters, $\mathcal{C}$ and $\mathcal{A}$ represent the full and active sets of constraints, respectively, and $p^{*}$ and $p_{\mathrm{init}}$ are the optimal and initial values of the optimization variables. ML-based end-to-end (direct) methods predict directly $p^{*}$ using regression, while hybrid (indirect) approaches predict either $p_{\mathrm{init}}$ (using regression) or $\mathcal{A}$ (using classification).}
    \label{fig:opf_interior_strategies}
\end{figure}

OPF problems are typically solved through interior-point methods~\cite{Fiacco68}, also known as barrier methods (Figure~\ref{fig:opf_interior_strategies}, left panel). 
One of the most widely used approaches is the primal-dual interior-point technique with a filter line-search \cite{Wachter06}.
These methods are robust but expensive, as they require the computation of the second derivative of the Lagrangian at each iteration.
Nevertheless, interior-point methods can be considered as baseline approaches to solving general OPF problems.

In order to reduce computational costs, various approximations are used.
The most typical approximation, called DC-OPF \cite{cain2012history}, makes the problem convex and reduces the number of variables and constraints.
Recent works apply the L-BFGS-B method~\cite{Tang17} or the coordinate-descent algorithm~\cite{Liu18} to get real time approximations of the AC-OPF problem.

A fruitful and new direction of research is to use machine learning (ML) techniques to solve operation and control problems for power grids.
For example, deep neural networks (DNN) have been deployed for grid state estimation and forecasting \cite{zhang2019real}, and reinforcement learning approaches to address the voltage control problem in distribution grids~\cite{yang2019two}.
Our focus in this paper is on the  ML approaches that are being deployed to predict the solution of OPF or Security-Constrained Unit Commitment problems~\cite{Xavier19}, shifting computational effort away from real-time to offline training. 
These black-box techniques roughly fall into two categories: regression and classification methods. 

The most widely used \emph{end-to-end} (or \emph{direct}) approaches try to predict the optimal OPF solution based on the grid parameters through regression techniques. 
As OPF is a constrained optimization problem, the solution is not a smooth function of the grid parameters, so properly training such regression models requires substantial amounts of data \cite{Guha19, Fioretto09}. 
There is also no guarantee that the solution satisfies all constraints, and violation of important constraints could lead to severe security issues for the grid.
Nevertheless, the predicted solution can instead be utilized as a starting point to initialize an interior-point method~\cite{Baker19} (Figure~\ref{fig:opf_interior_strategies}, middle panel).
This approach is often called a \emph{hybrid} (or \emph{indirect}) method as it combines ML-based technique with a subsequent OPF calculation.
Predicting a set-point close enough to the solution can significantly reduce the number of optimization iterations compared to the original problem \cite{Jamei19}, but the computational gain realized in practice is marginal for several reasons.
First, because only primal variables are initialized, the duals still need to converge, as interior-point methods require a minimum number of iterations even if the primals are set to their optimal values.
Trying to predict the duals as well makes the task even more challenging.
Second, if the initial values of primals are far from optimal, the optimization can lead to a different local minimum.
Finally, even if the predicted values are close to the optimal solution, as there are no guarantees on feasibility, this could locate in a region resulting in substantially longer solve times, or even convergence failure. 

To overcome this problem, one option is to attempt to obtain a feasible approximate solution without running a warm-start OPF optimization, but using cheaper post-processes instead. 
For instance, for DC-OPF Pan et al.~\cite{Pan19} use a DNN to map the load inputs to the outputs. 
However, instead of predicting the optimal value of all optimization variables, they predict only the active power of generators. 
Since there is a direct linear relationship between the phase angles and active powers via the admittance matrix, they compute the phase angles from the predicted active power by solving the corresponding linear system. 
Also, instead of directly predicting the active power, a simple linear transformation is predicted to automatically satisfy the corresponding minimum and maximum inequality constraints. 
Finally, since the prediction is still not necessarily a feasible solution, a projection of this prediction is applied that requires solving a quadratic program.
For AC-OPF Zamzam and Baker~\cite{Zamzam19} worked out an approach that has many similarities to the one described above. 
They also use a DNN to map the (active and reactive) loads to a subset of outputs and predict only the voltage magnitudes and active power outputs of generators. 
They also use a reparameterization of these variables so the boundary constraints of the predicted quantities are automatically satisfied. 
Finally, the voltage angles and reactive power outputs are obtained by solving the non-linear equations of the original OPF problem using the predicted quantities. 
Solving a nonlinear system is much faster than solving a nonlinear constrained optimization problem.
The above methods have been shown to have excellent results for small synthetic grids.
However, for larger grids one potential drawback is that these approaches might provide feasible but not necessarily optimal solutions.

Classification black-box methods leverage the observation that only a fraction of constraints is actually binding at the optimum~\cite{Zhou11}, so a reduced OPF problem can be formulated by keeping only the binding constraints.
Since this reduced problem still has the same objective function as the original, the solution should be equivalent to that of the original full problem (Figure~\ref{fig:opf_interior_strategies}, right panel).
This suggests a classification-based hybrid method, in which grid parameters are used to predict the binding status of each constraint and a reduced OPF problem is solved.
Deka and Misra~\cite{Misra18} identify all distinct active sets (i.e. all distinct structures of reduced OPFs) in the training data set and train a NN classifier (with cross-entropy loss function) to predict the corresponding active set given the load inputs.

However, classification methods can also lead to security issues through false negative predictions of the binding status of important constraints. 
By iteratively checking and adding violated constraints, and then solving the reduced OPF problem until all constraints of the full problem are satisfied, this issue can be avoided~\cite{Pineda20}. 
As the reduced OPF problem is much cheaper than the full one, this procedure (if converged in few iterations) can be very efficient.

This approach is compatible with current practices of some grid operators to solve OPF, where the transmission security constraints are enforced through an iterative procedure in which the solution at each iteration is checked against the base-case and \nmo{} contingency constraints:
all violated constraints are added to the model, and the procedure continues until no more violations are found~\cite{ma2009security}.
We hereafter refer to this approach as the \textit{iterative feasibility test}.

Focusing on the computational cost of obtaining an OPF solution by hybrid techniques suggests the use of an unconventional loss function that directly measures this cost (instead of addressing regression or classification errors). 
In~\cite{Jamei19} we combined a regression based hybrid approach with such an objective by minimizing the total number of the OPF solver iterations by predicting an appropriate warm-start for the interior-point primal variables.
Viewing the initialization as parameters of an OPF solver, we refer to this objective as a \textit{meta-loss} and its optimization as \textit{meta-optimization} since it is optimization of an optimizer.\footnote{Although there is overlap, we do not refer to this as meta-learning. A meta-learning framework would typically involve meta-training a single model over different data-sets (tasks) that is capable of few-shot learning on unseen data-sets at meta-test time. In contrast, here we have two models: an interior point solver that solves OPF problems, and a DNN that learns (input) parameters for the interior point solver for faster convergence. There is, however, no real notion of different tasks in this framework; training and testing all operate on the same grid.} 

Inspired by recent works in predicting active constraint sets \cite{Misra18, Deka19, Ng18} and using a computational cost based meta-loss objective function~\cite{Jamei19}, the main contributions of this paper are the following.
We introduce a classification-based hybrid method to cope with situations where:
1. the grid parameters have a wide distribution (i.e. they have a large deviation compared to the samples used in the above papers) resulting in a high number of distinct active sets.
2. the size of training data is limited compared to the space of potential active-sets. 
3. the solution guarantees feasibility and maintains optimality of the full problem achievable by the optimizer.
First, given the high number of distinct active sets, instead of predicting the appropriate set (as in~\cite{Misra18}), our method predicts the binding status of each inequality constraint and builds a reduced OPF model.
Second, in order to obtain a feasible solution we apply the iterative feasibility test.
It should be noted that the proposed meta-optimization procedure to predict the active sets of OPF is agnostic to the actual solver used to solve the OPF problem.
In this sense, the method can be used in combination with any appropriate solver beside the ones tested in this work.
Third, instead of using a conventional loss function, we introduce a meta-loss objective that measures the entire computational time of obtaining a solution.
The NN weights are optimized by minimizing this objective: we call this meta-optimization.
The meta-loss function can be applied universally for both regression and classification-based hybrid methods, which makes it possible to perform a thorough quantitative comparison of the two approaches. 
We demonstrate the capability of our method on several DC- and AC-OPF problems.
To understand the theoretical limits of hybrid approaches, we explore a perfect regressor and classifier, as guides for further research. 
Finally, the scalability of our method is investigated and a wide range of grid sizes are tested, some of which have not been explored previously due to computational cost.

In order to facilitate research reproducibility in the field, we have made the generated DC- and AC-OPF samples (\url{https://github.com/invenia/OPFSampler.jl}), as well as our code (\url{https://github.com/invenia/MetaOptOPF.jl}) publicly available.

\section{Methods}
\subsection{DC- and AC-OPF formulations}
In this work, we focus on the base-case OPF using DC- and AC-OPF formulations and do not consider any contingency scenarios.
Let $\mathbb{G} = (\mathcal{V}, \mathcal{E})$ define an directed graph with $\mathcal{V}$ set of buses and $\mathcal{E}$ set of directed edges such that $\mathcal{E} = \{(i, j)\ |\ i \to j;\ i, j \in \mathcal{V}\}$.
Also, let $\mathcal{G}$ and $\mathcal{G}_i$ denote the set of generators and the set of generators at bus $i$, respectively such that $\mathcal{G} = \bigcup \limits_{i \in \mathcal{V}} \mathcal{G}_i$; $\mathcal{V}_s$ the set of slack buses and $\mathcal{N}_i$ the set of buses adjacent to bus $i$.
Finally, let $L^p_i$ and $L^q_i$ denote the active and reactive power loads at bus $i$; $P_{g}$ and $Q_{g}$ the active and reactive power outputs of generator $g$ and $F^p_{ij}$ and ${F}^q_{ij}$ are the active and reactive power flows on line $(i, j)$, respectively. 
According to the implementation of these models in PowerModels.jl~\cite{Coffrin18} we used the following optimization problem for DC-OPF:
\begin{subequations}
\begin{align}
    \min \limits_{X^{\mathrm{DC}}}\ \sum \limits_{g \in \mathcal{G}} C_g(P_g)
\end{align}
subject to:
    \begin{align}
        \mathcal{C}_{\mathrm{eq}}
        &\begin{cases}
        &F^p_{ij} = \frac{\theta_{ij}}{x_{ij}},\hspace{1.9cm}\forall (i, j) \in \mathcal{E}\\
        &\theta_{i} = 0,\hspace{3.0cm}\forall i \in \mathcal{V}_s\\
        &\sum\limits_{j \in \mathcal{N}_i}F^p_{ij} = \sum\limits_{g \in \mathcal{G}_i} P_g - L^p_i,\hspace{0.35cm}\forall i \in \mathcal{V}
        \end{cases}
        \\
        \mathcal{C}_{\mathrm{ineq}}
        &\begin{cases}
        &\underline{P}_g \leq P_g \leq \overline{P}_g,\hspace{1.7cm}\forall g \in \mathcal{G}\\
       & |{F}^p_{ij}| \leq \overline{F}_{ij},\hspace{1.7cm}\forall (i, j) \in \mathcal{E}\\
       &|{\theta}_{ij}| \leq \overline{\theta}_{ij},\hspace{1.9cm}\forall (i, j) \in \mathcal{E}
       \end{cases}
    \end{align}
\end{subequations}
where $X^{\mathrm{DC}}$ denotes the vector of DC optimization variables, $C_{g}$ is the cost curve of generator $g$, $x_{ij}$ is the reactance of line $(i, j)$ and $\theta_{i}$ and $\theta_{ij} = \theta_{i} - \theta_{j}$ are the voltage angle at bus $i$ and voltage angle differences between bus $i$ and $j$, respecively. 
Lower and upper bounds are denoted by the corresponding underlined and overlined variables. 
The sets of equality ($\mathcal{C}_{\mathrm{eq}}$) and inequality ($\mathcal{C}_{\mathrm{ineq}}$) constraints are also indicated.

For AC-OPF, the following optimization model is considered:
\begin{subequations}
\begin{align}
    \min \limits_{X^{\mathrm{AC}}}\ \sum \limits_{g \in \mathcal{G}} C_g(P_g)
\end{align}
subject to:
    \begin{align}
        \mathcal{C}_{\mathrm{eq}}^{\mathrm{cvx}}
        &\begin{cases}
        &\sum\limits_{j \in \mathcal{N}_i}F^p_{ij} = \sum\limits_{g \in \mathcal{G}_i} P_g - L^p_i,\hspace{1.7cm}\forall i \in \mathcal{V}\\
        &\sum\limits_{j \in \mathcal{N}_i}\left(F^q_{ij} + \frac{b^c_{ij}}{2}v_i^2\right) = \sum\limits_{g \in \mathcal{G}_i} Q_g - L^q_i,\forall i \in \mathcal{V}\\
        &\theta_i = 0,\hspace{4.3cm}\forall i \in \mathcal{V}_s
        \end{cases}
        \\
        \mathcal{C}_{\mathrm{eq}}^{\mathrm{ncvx}}
        &\begin{cases} \label{eq.ncvx}
        &F^p_{ij} = \frac{r_{ij}}{r^2_{ij} + x^2_{ij}}\frac{v_i^2}{\tau_{ij}^2}-\frac{v_i v_j}{(r^2_{ij} + x^2_{ij})\tau_{ij}}\times\\
        &\left(r_{ij}\cos(\delta_{ij})-x_{ij}\sin(\delta_{ij})\right),\hspace{0.65cm}\forall (i, j) \in \mathcal{E} \\
        &F^q_{ij} = \left(\frac{-x_{ij}}{r^2_{ij} + x^2_{ij}}+\frac{b^c_{ij}}{2}\right)\frac{v_i^2}{\tau_{ij}^2}-\frac{v_i v_j}{(r^2_{ij} + x^2_{ij})\tau_{ij}}\times\\
        &\left(r_{ij}\sin(\delta_{ij})+x_{ij}\cos(\delta_{ij})\right),\hspace{0.65cm}\forall (i, j) \in \mathcal{E}\\
        &F^p_{ji} = \frac{r_{ij}}{r^2_{ij} + x^2_{ij}}v_j^2-\frac{v_i v_j}{(r^2_{ij} + x^2_{ij})\tau_{ij}}\times\\
        &(r_{ij}\cos(\delta_{ij})+x_{ij}\sin(\delta_{ij})),\hspace{0.7cm}\forall (i, j) \in \mathcal{E} \\
        &F^q_{ji} = \left(\frac{-x_{ij}}{r^2_{ij} + x^2_{ij}}+\frac{b^c_{ij}}{2}\right)v_j^2-\frac{v_i v_j}{(r^2_{ij} + x^2_{ij})\tau_{ij}}\times\\
        &\left(-r_{ij}\sin(\delta_{ij})+x_{ij}\cos(\delta_{ij})\right),\hspace{0.35cm}\forall (i, j) \in \mathcal{E}
        \end{cases}
        \\
        \mathcal{C}_{\mathrm{ineq}}
        &\begin{cases}
        &\underline{v}_i \leq v_i \leq \overline{v}_i,\hspace{3.4cm}\forall i \in \mathcal{V}\\
        &\underline{P}_g \leq P_g \leq \overline{P}_g,\hspace{2.95cm}\forall g \in \mathcal{G}\\
        &\underline{Q}_g \leq Q_g \leq \overline{Q}_g,\hspace{2.9cm}\forall g \in \mathcal{G}\\
       &|{F}^p_{ij}| \leq \overline{F}_{ij},\hspace{3.0cm}\forall (i, j) \in \mathcal{E}\\
       &|{F}^q_{ij}| \leq \overline{F}_{ij},\hspace{3.0cm}\forall (i, j) \in \mathcal{E}\\
       &|{\theta}_{ij}| \leq \overline{\theta}_{ij},\hspace{3.15cm}\forall (i, j) \in \mathcal{E}\\
       &{{F}^p_{ij}}^2 + {{F}^q_{ij}}^2\leq {\overline{F}_{ij}}^2,\hspace{1.8cm}\forall (i, j) \in \mathcal{E}
        \end{cases}
    \end{align}
\end{subequations}
where, beside the previous definitions, $X^{\mathrm{AC}}$ is the vector of AC optimization variables, $v_i$ denotes the magnitude of voltage at bus $i$, and $r_{ij}$ and $b^c_{ij}$ are the resistance and shunt charging susceptance of line $(i, j)$, respectively.
$\tau_{ij}$ and $\theta^t_{ij}$ are the magnitude and angle of the phase shifter tap ratio on line $(i, j)$ and for brevity we defined $\delta_{ij} = {\theta}_{ij} - \theta^t_{ij}$.
If there is no phase shifter on the line, then $\tau_{ij} = 1$ and $\theta^t_{ij} = 0$.
For AC-OPF, three types of constraints are distinguished: convex equality ($\mathcal{C}_{\mathrm{eq}}^{\mathrm{cvx}}$), non-convex equality ($\mathcal{C}_{\mathrm{eq}}^{\mathrm{ncvx}}$), and inequality ($\mathcal{C}_{\mathrm{ineq}}$) sets.

\subsection{Meta-optimization for Regression-based Hybrid Approaches}
In order to introduce the concept of a meta-loss as an alternative objective function, we briefly describe our previous regression-based model~\cite{Jamei19}. 

Conventional supervised regression techniques typically use loss functions based on a distance between the training ground-truth and predicted output value, such as mean squared error or mean absolute error~\cite{Bishop06}. 
In general, each dimension of the target variable is treated equally in the loss function.
However, the shape of the Lagrangian landscape of the OPF problem as a function of the optimization variables is far from isotropic~\cite{Mones18}, implying that optimization under such an objective does not necessarily minimize the warm-started OPF solution time.
The reason is that trying to derive initial values for optimization variables using empirical risk minimization techniques cannot guarantee feasibility, despite the accuracy of the prediction to the ground truth. 
Interior-point methods start by first moving the system into a feasible region, thereby potentially altering the initial position significantly. 
Consequently, warm-starting from an infeasible point can be inefficient.

\begin{figure}[H]
    \centerline{\includegraphics[width=1.0\textwidth]{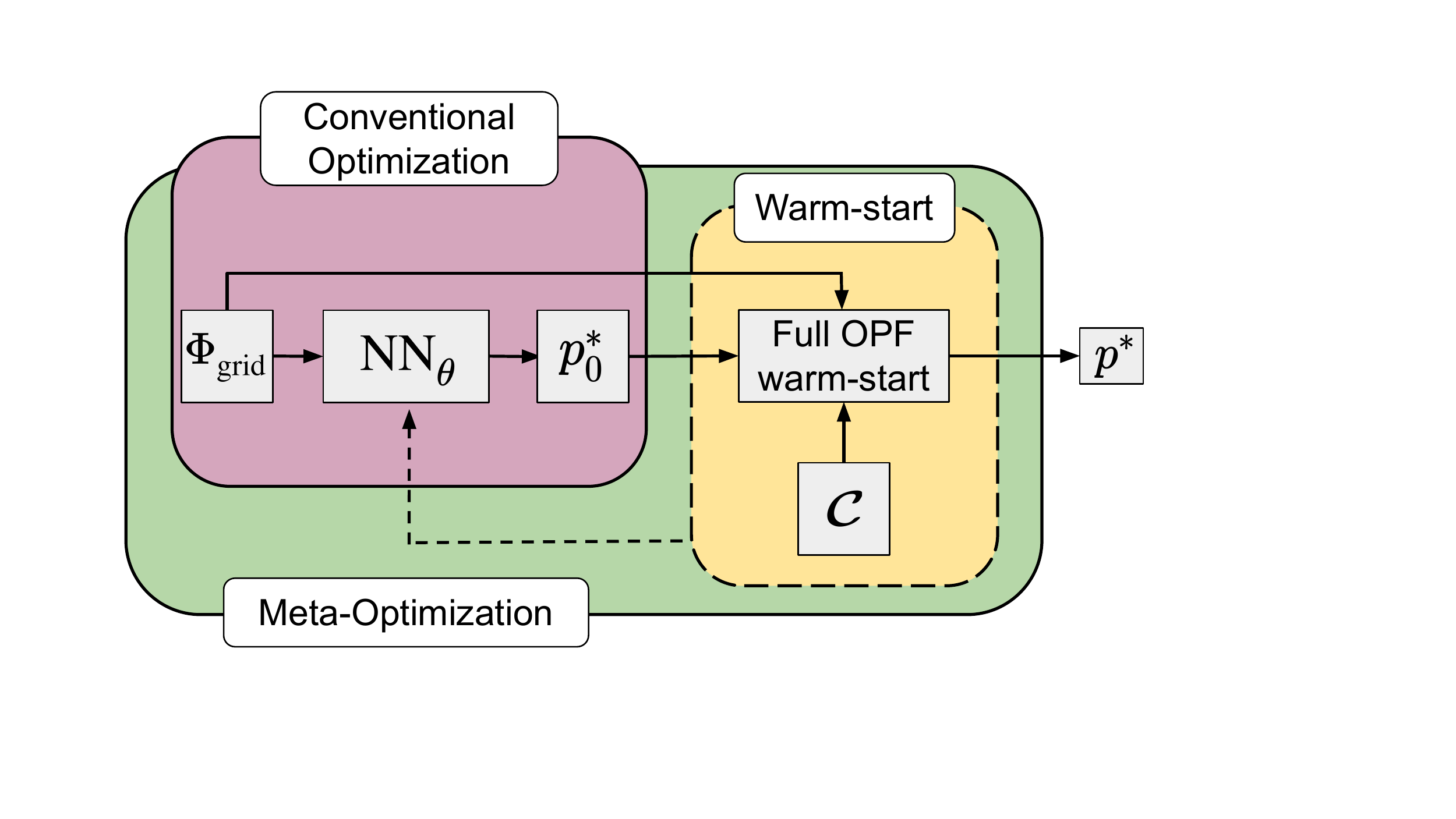}}
    \caption{Flowchart of the meta-optimization procedure using a NN regressor with warm-start. The initial values of weights $\theta$ for meta-optimization of the meta-loss are obtained from conventional training with a regression loss. $\Phi_{\mathrm{grid}}$ is the vector of grid parameters, NN$_{\theta}$ represents the regressor with weights $\theta$. The meta-loss is computed as the solve time or the total number of optimization steps of the warm-started OPF. $p_{0}^{*}$ is the initial value of the optimization variables and $\mathcal{C}$ denotes the full set of constraints of the problem.}
    \label{fig:regression_meta}
\end{figure}

Instead, we proposed a meta-loss function that directly measures the computational cost of solving the (warm-started) OPF problem~\cite{Jamei19}.
One measure of the computational cost can be defined by the number of iterations required to reach the optimal solution from the initialization point.
This is a deterministic and noise-free measure of the computational cost. 
Since the warm-started OPF has exactly the same formulation as the original OPF problem, the comparative number of iterations represents the improvement in computational cost.
We applied a neural network (NN), with parameters determined by minimizing the meta-loss function (meta-optimization) on the training set  (Figure~\ref{fig:regression_meta}).
As this meta-loss is a non-differentiable function with respect to the NN weights, back-propagation cannot be used.
As an alternative, we employed the Particle Swarm Optimization (PSO)~\cite{Kennedy95} to find an optimal meta-loss in the NN weight space.
PSO is a gradient-free meta-heuristic algorithm inspired by the concept of swarm intelligence that can be found in nature among certain animal groups.
The method applies a set of particles ($N_{\mathrm{p}}$), with the particle dynamics at each optimization step influenced by both the individual (best position found by the particle) and collective (best position found among all particles) knowledge.
Since each optimization step of PSO requires $N_{\mathrm{p}}$ computation of the meta-loss function, in this work we used its adaptive version~\cite{Zhan09} that has an improved convergence rate.
We also note that although PSO was originally introduced as a global optimization technique due to the high dimensionality of the weight space, we use it here as a local optimization technique.
Therefore, the particles were initiated with a small random perturbation from a position provided by the optimal weights for a conventional loss function~\cite{modpso}.

At each step of meta-optimization, the PSO particles try to optimize the meta-loss by varying the NN weights, predict the initial values of the optimization variables, and solve the corresponding warm-start OPFs of the training data. 
This requires solving OPF multiple times (as the predictor changes) leading to the following computational cost: with $N_{\mathrm{t}}$ meta-training samples, $N_{\mathrm{p}}$ PSO particles and $N_{\mathrm{s}}$ meta-optimization steps, $N_{\mathrm{t}} \times N_{\mathrm{p}} \times N_{\mathrm{s}}$ full OPF problems with warm-start must be solved. 
However, it is a highly parallelizable problem among the PSO particles.
Also, it is straightforward to start the meta-optimization from a pre-trained NN, under a conventional regression loss. 
We demonstrated the capability of this meta-optimization for two synthetic grids using DC-OPF problems \cite{Jamei19}.

\subsection{Meta-optimization for Classification-based Hybrid Approaches}
\begin{figure}[H]
    \centerline{\includegraphics[width=1.0\textwidth]{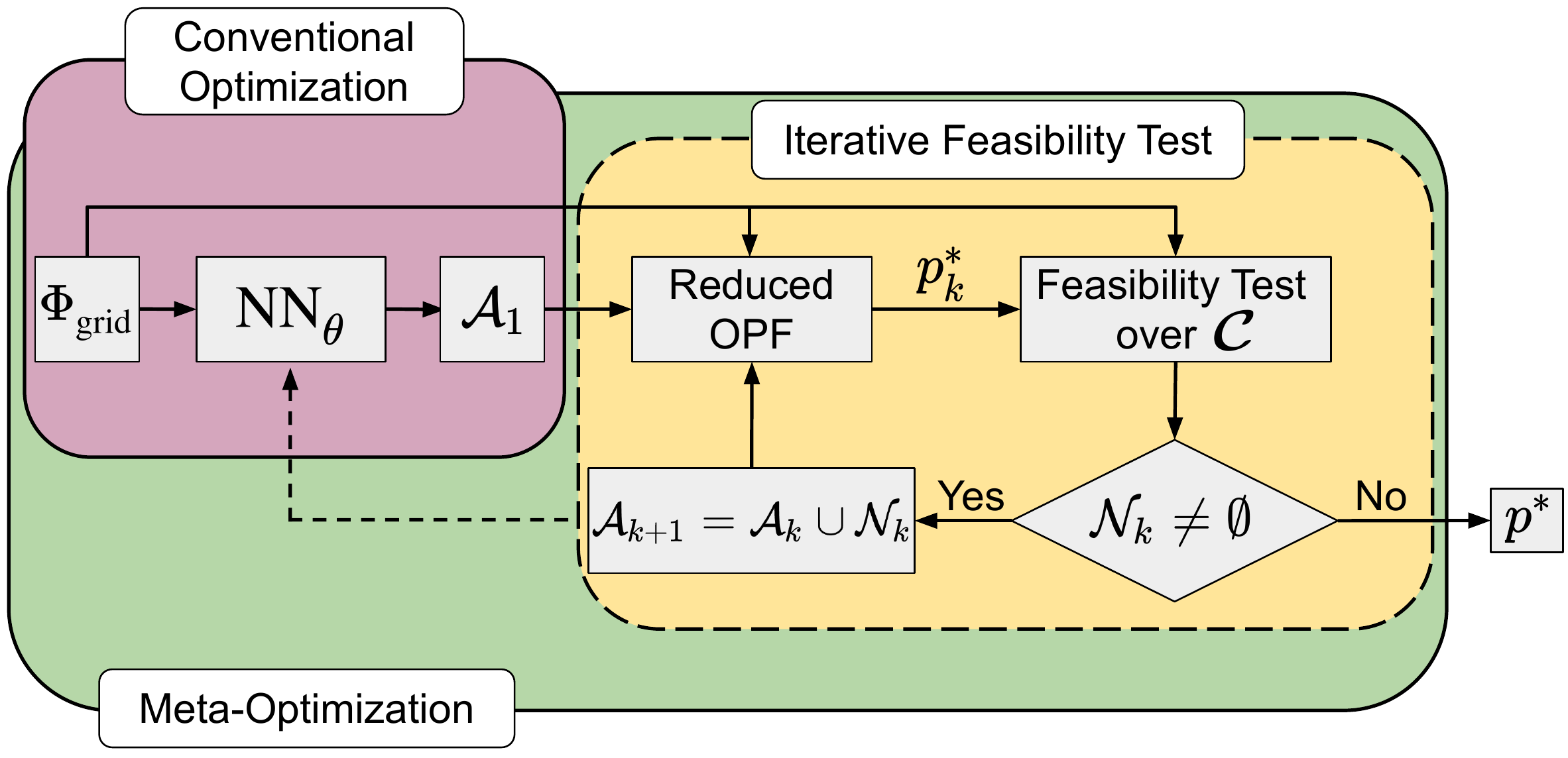}}
    \caption{Flowchart of meta-optimization using a NN classifier under the feasibility iteration procedure. Conventional optimization of a classification loss, which provides initial weights $\theta$, is followed by a meta-optimization of the meta-loss. $\Phi_{\mathrm{grid}}$ is the vector of grid parameters, NN$_{\theta}$ represents the classifier with weights $\theta$. The meta-loss is computed within the iterative feasibility test, where $\mathcal{C}$ denotes the full set of constraints of the original OPF problem, $\mathcal{A}_{k}$ is the actual set used in the reduced problem and $\mathcal{N}_{k}$ is the set of violated constraints. $p^{*}_{k}$ is the solution of the corresponding reduced problem, where $k = 1 \dots K$ is the iteration index. The final solution $p_{K}^{*} = p^{*}$ at $k=K$ is obtained when $\mathcal{N}_K = \emptyset$.}
    \label{fig:classification_meta}
\end{figure}

The first step of our new hybrid method is to train a NN-based classifier using grid parameters as features to predict the binding status of the constraints of the full OPF problem.
A reduced OPF problem has the same objective function as the full problem, but only retains those constraints that were predicted to be binding by the classifier. As there may be violated constraints not included in the reduced model, we use the \textit{iterative feasibility test} to ensure convergence to an optimal solution of the full problem. 
The procedure has the following steps (Figure~\ref{fig:classification_meta}): 
\begin{enumerate}
    \item An initial reduced set of constraints $\mathcal{A}_1$ is proposed by the classifier. A solution $p^{*}_1$ is then obtained by solving the reduced problem.
    \item In each feasibility iteration, $k \in 1 \dots K$, the solution $p^{*}_k$ of the reduced problem is validated against the constraints $\mathcal{C}$ of the original full formulation.
    \item At each step $k$, the violated constraints $\mathcal{N}_k$ are added to the set of considered constraints to form $\mathcal{A}_{k+1}$ $=$ $\mathcal{A}_{k} \cup \mathcal{N}_k$.\label{step:vcadd}
    \item This procedure repeats until no violations are found (i.e.~$\mathcal{N}_K$ $=$ $\emptyset$), and the solution $p^{*}_K$ satisfies all original constraints $\mathcal{C}$.
    At this point, we have found a solution to the full problem ($p^{*}$).
\end{enumerate}

The goal is to find NN weights that minimize the total computational time of the iterative feasibility test.
However, as we will demonstrate, minimizing a cross-entropy loss function to obtain such weights is not straightforward.
First, the number of cycles in the iterative procedure described above is much more sensitive to the false negative than false positive predictions of the binding status.
Second, different constraints can be more or less important depending on the actual congestion regime and binding status.
These suggest the use of a more sophisticated objective function, for instance a weighted cross-entropy with appropriate weights for the corresponding terms.
The weights as hyper-parameters then can be optimized to achieve an objective that can adapt to the above requirements.
However, an alternative objective can be defined as the total computational time of the iterative feasibility test procedure.
Since this directly measures the performance of a sequence of reduced OPF optimizations, we call it a meta-loss function and its optimization over a training data set as meta-optimization.

The meta-loss objective, therefore, includes the solution time of a sequence of reduced OPF problems. 
Similarly to the meta-loss defined for the regression approach, it measures the computational cost of obtaining a solution of the full problem and unlike weighted cross-entropy it does not require additional hyper-parameters to be optimized.
As the meta-loss is a non-differentiable function of the classifier weights, we optimize it using the gradient-free PSO method.
At each step of the meta-optimizations, each PSO particle varies the NN weights, predicts the binding status of the constraints, and performs the iterative feasibility test of the training data to optimize the meta-loss.
The meta-optimization has therefore the following computational cost: with $N_{\mathrm{t}}$ meta-training examples, $N_{\mathrm{p}}$ particles, and $N_{\mathrm{s}}$ meta-optimization steps, $\sum_{i=1}^{N} K_{i}$ reduced OPF calculations are performed, where $N = N_{\mathrm{t}} \times N_{\mathrm{p}} \times N_{\mathrm{s}}$, and $K_{i}$ is the number of feasibility test iterations of the $i$th reduced OPF problem.

Of all these parameters, $N_{\mathrm{t}}$, $N_{\mathrm{p}}$ and $N_{\mathrm{s}}$ are the hyperparameters we control. 
The values $\left\{ K_{i} \right\}_{i=1}^{N}$, however, are dependent upon the classifier performance. 
In our experience, the procedure usually converges within a few iterations to the solution of the full problem (typically $1$--$10$ for tested grids). 
Further, we note that instead of just extending the previous set of active constraints with violations, an alternative to step~\ref{step:vcadd} would be to also discard constraints that were found to be non-binding in $\mathcal{A}_k$. 
This alternative approach can theoretically lead to infinite loops when competing constraints switch their binding status from one to another between consecutive iterations.
We also found it to have a slower convergence behavior in practice than the extension-only version we recommend in step~\ref{step:vcadd}.

To reduce the required number of steps of meta-optimization, we initialize the NN classifier by training under a conventional objective for classification.
As discussed previously for regression, optimizing such an objective does not necessarily minimize the computational cost of obtaining a solution of the full problem. 
In practice, however, we achieve reasonable results by training with a cheap surrogate objective (conventional loss) first, followed by training under the more expensive meta-loss objective.   
We summarize the differences between regression- and classification-based approaches in Table~\ref{tab:meta_models}.

\begin{table}[H]
\small
\caption{Comparison of regression- and classification-based hybrid approaches using meta-optimization.}
\label{tab:meta_models}
\vskip 0.1in
\centering
\resizebox{\columnwidth}{!}{
    \begin{tabular}{lll}
    \toprule
    Property & Regression & Classification \\
    \midrule
    Input & $\Phi_{\mathrm{grid}}$ & $\Phi_{\mathrm{grid}}$ \\
    Output & $p_{0}^{*}$ & $\mathcal{A}_{1}$ \\
    OPF problem type to solve & full OPF with warm-start & reduced OPF formulations \\
    Meta-loss & solve time (or \# iterations) & total solve time \\
    Meta-optimizer & PSO varying NN weights & PSO varying NN weights \\
    Cost of meta-optimization & $N_{\mathrm{t}} \times N_{\mathrm{p}} \times N_{\mathrm{s}}$ & $\sum \limits_{i=1}^{N_{\mathrm{t}} \times N_{\mathrm{p}} \times N_{\mathrm{s}}} K_{i}$ \\
    \bottomrule
    \end{tabular}
}
\vskip -0.1in
\end{table}

\section{Experimental Analysis}

\subsection{OPF Framework}
Several synthetic grids from the Power Grid Library \cite{Baba19} were used.
DC- and AC-OPF models were solved within the PowerModels.jl \cite{Coffrin18} OPF package written in Julia \cite{bezanson2012julia}.
For interior-point methods, we used the Ipopt \cite{Wachter06} solver.

\subsection{Input Sample Generation}
\label{sec:sample_gen}
In order to explore a variety of distinct active sets of constraints for the synthetic cases and mimic the time-varying behavior of the OPF input parameters, grid parameter samples with feasible OPF solutions were generated by varying their original values in the grid data-set.
In particular, for each grid $10$k DC-OPF samples were produced by rescaling each nodal load active power by factors independently drawn from uniform distribution of the form ~$\mathcal{U}(0.85, 1.15)$, and rescaling each maximum active power output of generators, line thermal ratings and line reactance values by scaling factors drawn from ~$\mathcal{U}(0.9, 1.1)$. 
Accordingly, the input parameter vector for DC-OPF cases was defined as $\Phi_{\mathrm{grid}}^{\mathrm{DC}} = \{L_i^p, \overline{P}_g, \overline{F}_{ij}, x_{ij}\}$.

For AC-OPF, $1$k samples were generated for the studied synthetic grids. 
Beside the parameters that were changed for DC-OPF, rescaled nodal load reactive power, maximum reactive power output of generators, and line resistance values were produced by scaling factors sampled from ~$\mathcal{U}(0.9, 1.1)$.
Therefore, for AC-OPF cases the input parameter vector consisted of the following parameters: $\Phi_{\mathrm{grid}}^{\mathrm{AC}} = \{L_i^p, L_i^q, \overline{P}_g, \overline{Q}_g, \overline{F}_{ij}, x_{ij}, r_{ij}\}$.

\subsection{Technical Details of the Model}
\label{sec:technical_details}
\subsubsection{NN Architecture}
Each constraint was predicted to be binding or non-binding by a multi-label classifier. 
Correspondingly, a binary cross-entropy loss was used with the following architecture, in the Julia Flux.jl package \cite{innes2018flux}.
Two fully connected hidden layers were each followed by a BatchNorm layer \cite{Ioffe15} and a ReLU activation function \cite{Vinod2010ICML}. 
A Dropout layer \cite{Srivastava2014JMLR} with a dropout fraction of 0.4 was added after each BatchNorm layer. 
The final output layer had a sigmoid activation function. 
The input and output sizes of the NN were determined by the number of grid parameters and the cardinality of all inequality constraints, respectively (see Tables~\ref{tab:active_sets_dc} and~\ref{tab:active_sets_ac} for details), while the middle layer size was $50 \times 50$.

\subsubsection{Conventional Optimization}
Samples were split randomly into training, validation, and test sets of 70\%, 20\% and 10\%, respectively.
Hereafter, when referring to $10$k of DC-OPF or $1$k of AC-OPF samples, we refer to the total data set split as such. 
Mini-batch sizes of $10$ and $100$ were used with $1$k and $10$k samples, respectively.
Training was carried out using the ADAM optimizer \cite{kingma2014adam} (with learning-rate $\eta = 10^{-4}$ and parameters $\beta_{1} = 0.9$ and $\beta_{2} = 0.999$), using early stopping with a patience of $10$ determined on a validation set after a $50$ epoch burn-in period. 

\subsubsection{Binding Status of Constraints}
As the power flow equality constraints are always binding, we limited the binding-state prediction to inequality constraints only. 
The binding status of the lower-bound constraints of generator output power was not predicted (but force-set to be always binding) as the reduced OPF problem may become unbounded with their removal.
For similar optimization stability reasons, for AC-OPF, the binding status of lower and upper bound limits of voltage magnitudes were not predicted either, and always set to binding.
Therefore, the following inequality constraints were predicted: upper limit of generator active outputs, lower/upper limits of real power flows and voltage angle differences for DC-OPF and upper limit of generator active outputs, lower/upper limits of generator reactive outputs, active and reactive power flows, voltage angle differences and upper limits of the squared of complex power flows for AC-OPF.
The binding status of the constraints was assigned by checking each side of the inequality constraints.
We considered a constraint binding if either it was violated or the absolute value of the difference between the two sides was less than a fixed threshold value set at $10^{-5}$.

\subsubsection{Meta-optimization}\label{sec:method_metaoptimization}
During meta-optimization, the NN weights obtained from conventional optimization were further varied to optimize the meta-loss objective, defined as the total computational time to solve each OPF problem in the meta-training data. 
At each evaluation, the meta-training data was randomly sub-sampled from the training data with $N_{\mathrm{t}} = 100$ that improved the model to avoid overfitting. 
For each investigated grid, $10$ particles and $50$ iterations of PSO were run, using the Optim.jl package~\cite{mogensen2018optim}. 
The package was slightly modified to improve the particle initialization.
The starting position of the particles in the NN weight space was derived from the weights of the conventionally optimized NN and each component was perturbed by a random number drawn from a normal distribution with zero mean and standard deviation set at the absolute value of the component. 

Finally, in order to avoid converging to trivial minima of the meta-loss (discussed in Section-\ref{sec:improving_metaloss}), a penalty term was introduced during meta-optimization: if the number of predicted active constraints was higher than a threshold defined as twice the average number of the active constraints in the training data, the value of the meta-loss function was set to infinity.

\subsection{Computing Resources}
Shorter analyses with absolute computational times were run on Macbook Pro machines (2.9 GHz Quad-Core Intel Core i7 processor for Tables~\ref{tab:active_sets_dc} and~\ref{tab:active_sets_ac}; and 3.5 GHz Quad-Core Intel Core i7 processor for Figures~\ref{fig:constraints_vs_cost} and~\ref{fig:loss_vs_metaloss}, respectively.)
Meta-optimization experiments were carried out by using various machines on Amazon Elastic Compute Cloud.

\section{Results}
\subsection{Distinct Active Sets in DC- and AC-OPF Samples}
Based on the generated samples, we first investigated the number of unique active sets (congestion regimes) of several synthetic grids.
Table~\ref{tab:active_sets_dc} shows the results for DC-OPF samples with $10$k and a random $1$k subset. 
It also provides the number of grid parameters $\dim(\Phi_{\mathrm{grid}})$ (i.e., the classifier input size), the number of inequality constraints ($|\mathcal{C}_{\mathrm{ineq}}|$), where the binding status is predicted (the classifier output size) and also the number of equality constraints ($|\mathcal{C}_{\mathrm{eq}}|$) that are always binding.

For the $1$k subset, we compared the number of distinct active sets to those reported in \cite{Ng18}, which were generated by scaling nodal load with a factor drawn from a normal distribution with $\mu=1.0$ and  $\sigma=0.03$.
In the data presented here, the number of unique active sets is generally significantly higher that can be attributed to two major intentional differences: 1) varying more parameters beyond load, and 2) selecting a wider deviation for the load scaling values.

It is also clear from this setup that a sample size of $1$k is too limited to cover all possible distinct active sets for these grids. 
When extending the number of samples to $10$k we observe a further increase in the number of active sets. 
For larger grids, this is capped at the number of samples meaning that every sample has a unique active set. 
This indicates that under the sampling distribution of grid parameters, convergence to the real distribution of active-sets becomes increasingly poor, particularly for the larger grids with realistic sampling numbers.

\begin{table}[H] 
\small
\caption{Grid characteristics and number of unique active sets for different DC-OPF cases, using $1$K and $10$K samples.}
\label{tab:active_sets_dc}
\def\na{---}
\centering
    \resizebox{0.95\columnwidth}{!}{
    \begin{tabular}{lrrrrrr}
    \toprule
    \multirow{3}{*}{Case} & \multirow{3}{*}{$\dim(\Phi_{\mathrm{grid}})$} & \multirow{3}{*}{$|\mathcal{C}_{\mathrm{ineq}}|$} & \multirow{3}{*}{$|\mathcal{C}_{\mathrm{eq}}|$} & \multicolumn{3}{c}{Number of active sets} \\
    \cmidrule(r){5-7}
    & & & & {Ref~\cite{Ng18}} & \multicolumn{2}{c}{This work} \\
    \cmidrule(r){5-7}
    & & & & $1$k & $1$k & $10$k \\
     \midrule
     {24-ieee-rts} & 125 & 208 & 63 & 5 & 15 & 18 \\
     {30-ieee} & 105 & 168 & 72 & 1 & 1 & 1  \\
     {39-epri} & 123 & 204 & 86 & 2 & 8 & 12 \\
     {57-ieee} & 206 & 324 & 138 & 3 & 8 & 9 \\
     {73-ieee-rts} & 387 & 648 & 194 & 21 & 8 & 48 \\
     {118-ieee} & 490 & 768 & 305 & 2 & 66 & 122 \\
     {162-ieee-dtc} & 693 & 1152 & 447 & 9 & 188 & 513 \\
     {300-ieee} & 1080 & 1754 & 712 & 22 & 835 & 5145 \\
     {588-sdet} & 1846 & 2916 & 1275 & \na & 826 & 5004 \\
     {1354-pegase} & 4915 & 7922 & 3346 & \na & 997 & 9506 \\
     {2853-sdet} & 10275 & 16750 & 6775 & \na & 1000 & 10000 \\
     {4661-sdet} & 15401 & 24944 & 10659 & \na & 1000 & 10000 \\
     {9241-pegase} & 38438 & 63402 & 25291 & \na & 1000 & 10000 \\
     \bottomrule
    \end{tabular}
    }
\end{table} 
We performed a similar analysis of grid properties for AC-OPF cases using $1$k samples (Table~\ref{tab:active_sets_ac}).
As expected, the number of grid parameters and number of constraints are significantly higher than those of the corresponding DC-OPF cases.
In Table~\ref{tab:active_sets_ac}, we split the number of equality constraints---that are always binding---into two sets: convex and non-convex.
The number of convex equality constraints ($|\mathcal{C}_{\mathrm{eq}}^{\mathrm{cvx}}|$) is very similar to those of DC-OPF, but there is also a great number of the non-convex equality constraints ($|\mathcal{C}_{\mathrm{eq}}^{\mathrm{ncvx}}|$).
We note that the computation of the non-convex equality constraints and their first and second derivatives is the most expensive part of an interior-point optimization.
Given that the systems are larger, it is not surprising that the number of distinct active sets is higher than those of the corresponding DC-OPF cases.

For both formulations, the exponentially increasing number of distinct congestion regimes suggests that it would be more efficient to predict the binding status of individual constraints, rather than predicting the active set directly.
\begin{table}[H]
\small
\caption{Grid characteristics and number of unique active sets for different AC-OPF cases, using $1$K samples.}
\label{tab:active_sets_ac}
\centering
    \resizebox{\columnwidth}{!}{
    \begin{tabular}{lrrrrr}
    \toprule
    Case & $\dim(\Phi_{\mathrm{grid}})$ & $|\mathcal{C}_{\mathrm{ineq}}|$ & $|\mathcal{C}_{\mathrm{eq}}^{\mathrm{cvx}}|$ & $|\mathcal{C}_{\mathrm{eq}}^{\mathrm{ncvx}}|$ & Number of active sets ($1$k) \\
     \midrule
     {24-ieee-rts} & 214 & 628 & 49 & 152 & 39 \\
     {30-ieee} & 177 & 576 & 61 & 164 & 8 \\
     {39-epri} & 200 & 670 & 79 & 184 & 154 \\
     {57-ieee} & 338 & 1098 & 115 & 320 & 7 \\
     {73-ieee-rts} & 660 & 1958 & 147 & 480 & 523 \\
     {118-ieee} & 864 & 2670 & 237 & 744 & 799 \\
     {162-ieee-dtc} & 1102 & 3772 & 325 & 1136 & 812 \\
     {300-ieee} & 1773 & 5804 & 601 & 1644 & 1000 \\
     {588-sdet} & 3006 & 9770 & 1177 & 2744 & 1000 \\
     {1354-pegase} & 7839 & 27078 & 2709 & 7964 & 1000 \\
     {2853-sdet} & 16629 & 55462 & 5707 & 15684 & 1000 \\
     \bottomrule
    \end{tabular}
    }
\end{table}

\subsection{Maximum Achievable Gains}
To compare the utility of a regression or classification approach, we begin with the estimation of the expected empirical limit of the achievable computational gain for different solvers.
In this setup, we explicitly refer to the gain achievable whilst keeping the feasibility guarantees of the solvers, which is not comparable with methods that drop the (expensive) feasibility guarantee.  
In practice, this is equivalent to computing the gain of computational cost of the perfect regressor or classifier in the corresponding framework of hybrid approaches. 
The perfect regressor and classifier are hypothetical 'perfect' predictors.
Instead of a trained model, for a perfect regressor-based hybrid model, we warm-start an OPF solver using the values of the primal variables at the solution.
Similarly, for a perfect classification-based hybrid model, we solve a perfectly reduced OPF problem with only the active set of binding constraints in the reformulation.
Therefore, we compute the average maximum achievable gain for several grids using DC- and AC-OPF formulations with $1$k samples.
We define the gain of the computational cost to the full OPF problem as: 
\begin{equation}
\label{eqn:gain}
\mathrm{Gain}(t_\mathrm{ML}) = 100\frac{t_{\mathrm{f}} - t_{\mathrm{ML}}}{t_\mathrm{f}}
\end{equation} 
where $t_{\mathrm{f}}$ and $t_{\mathrm{ML}}$ are the computational times of the original full OPF problem and the specific machine learning based approach, respectively.
Throughout, we refer to the computational solve-time, either full, reduced, or warm-started as the meta-loss.
Here, we evaluate the average of $\mathrm{Gain}(t_\mathrm{ML}^{*}) \ge \mathrm{Gain}(t_\mathrm{ML})$, where $t_\mathrm{ML}^{*}$ is the computational time of the corresponding perfect predictor-based hybrid approaches.

Among the interior-point solvers we used, only Ipopt had warm-start capability, limited to primal variables only, and for consistency, the maximum achievable gain for both regression and classification was investigated with this solver, where the value of its \texttt{bound\_push} and \texttt{bound\_frac} parameters were set to $10^{-9}$ for warm-start optimizations.
For each sample, we compared the optimal value of the objective of the warm-start and reduced OPF formulations to the solution of the full problem and found that they were indeed equal.
This was especially necessary for AC-OPF cases, where finding the same solution is less evident due to the non-convex nature of the problem.

Table~\ref{tab:mag_acdc} presents the results for both formulations.
In the case of the DC formulation, we observe that the maximum achievable gain of the regression-based hybrid approach is in general somewhat lower than that of the classification approach, especially for larger grids.
Further, while the maximal gain for regression shows little correlation with the grid size, there is a much stronger correlation for classification, indicating a better scaling when moving to larger grids.

Given the DC-OPF is a linear problem we can draw some qualitative conclusions regarding the system size and gain.
In the case of perfect regression, the size of the optimization problem is equal to the original OPF problem and the gain is determined by the convergence of dual variables that does not seem to depend on the size.
However, the gain of a perfect classifier-based hybrid method is primarily governed by the size of the reduced OPF problem compared to the full problem and it roughly depends on the ratio of the number of inequality constraints and the number of all constraints of the full OPF formulation (assuming that only a fraction of inequality constraints is actually active).
\begin{table}[H]
\small
\caption{Maximum achievable gains of warm-start with primal variables (perfect regression-based hybrid model) and reduced OPF formulations (perfect classification-based hybrid model) methods for several grids using DC- and AC-OPF formulations and the Ipopt solver.}
\label{tab:mag_acdc}
\def\na{---}
\centering
    \resizebox{\columnwidth}{!}{
    \begin{tabular}{lrr|rr}
    \toprule
    \multirow{2}{*}{Case} & \multicolumn{2}{c|}{DC Gain (\%)} & \multicolumn{2}{c}{AC Gain (\%)}\\
    \cmidrule(r){2-5}
    & Regression & Classification & Regression & Classification \\
     \midrule
     {24-ieee-rts} & $30.9\pm0.7$  &  $29.9\pm0.7$  & $27.0\pm0.6$  &  $25.2\pm0.6$ \\
     {30-ieee} & $33.9\pm0.5$  &  $28.3\pm0.5$  & $7.9\pm0.8$  &  $32.0\pm0.9$ \\
     {39-epri} & $52.7\pm0.4$  &  $28.0\pm0.4$  & $46.0\pm0.6$  &  $29.7\pm0.6$ \\
     {57-ieee} & $27.1\pm0.6$  &  $38.8\pm0.3$  &  $21.4\pm0.7$  &  $30.6\pm0.7$ \\
     {73-ieee-rts} & $29.7\pm0.3$  &  $36.8\pm0.3$  & $33.5\pm0.7$  &  $27.6\pm0.5$ \\
     {118-ieee} & $22.4\pm0.5$  &  $47.6\pm0.4$  & $15.8\pm0.6$  &  $31.1\pm0.4$ \\
     {162-ieee-dtc} & $55.4\pm0.4$  &  $47.3\pm0.3$  & $40.4\pm1.0$  &  $21.9\pm0.7$ \\
     {300-ieee} & $44.1\pm0.4$  &  $45.7\pm0.3$  & $37.2\pm1.4$  &  $17.4\pm0.6$ \\
     {588-sdet} & $28.5\pm0.5$  &  $57.0\pm0.3$  & $-18.3\pm1.0$  &  $12.2\pm0.8$ \\
     {1354-pegase} & $47.6\pm0.4$  &  $47.0\pm0.4$  & $1.6\pm1.3$  &  $35.1\pm0.4$ \\
     {2853-sdet} & $34.8\pm0.3$  &  $54.6\pm0.2$  & $-9.9\pm0.5$ &  $27.4\pm0.3$ \\
     {4661-sdet} & $38.0\pm0.3$  &  $45.1\pm0.3$  & \na & \na\\
     {9241-pegase} & $40.2\pm0.6$  &  $52.7\pm0.6$  & \na & \na \\
     \bottomrule
    \end{tabular}
    }
\end{table} 

For AC-OPF, we found that the maximum achievable gain is more moderate for both regression and classification compared to those of DC-OPF.
With the AC-OPF formulation, the gain of the perfect regression-based hybrid approach did not show a correlation with the system size, and for some cases we observed even negative gain with warm-started OPF.
Unlike the DC-OPF case, the gain of the perfect classification cannot be related simply to the ratio of inequality and equality constraints anymore: the computationally most expensive part is the calculation of the first and second derivatives of the non-convex equality constraints that are always binding (see Table~\ref{tab:active_sets_ac}).
In conclusion, we found that for DC-OPF the maximum achievable gain is significantly larger for classification ($\approx 50\%$) than for regression (at least for larger grid sizes).
No correlation was found between the system size and the gain of perfect regression, while a weak correlation was observed between the grid size and the gain of perfect classification.
For AC-OPF, the maximum achievable gains are significantly lower than those for DC-OPF but for larger grids, classification can still provide some improvement.

Finally, for DC formulation we computed the gains of the perfect classifiers using two other convex solvers: ECOS \cite{domahidi2013ecos} and OSQP \cite{osqp}.
We found similar trends of the maximal gains to those of Ipopt and for some cases these solvers slightly outperformed Ipopt.
However, we note that our meta-optimization method and the corresponding gain it provides is agnostic to the applied solver and therefore for the rest of this work we use the Ipopt solver to be consistent between AC and DC formulations and also for other practical considerations.

\subsection{Meta-loss as a Function of False Negative and False Positive Predictions}
We extended the empirical investigations away from perfect performance and examined the asymmetric effect of error in binding-constraint classification.
Specifically, we investigated the effect of increasing false negative (i.e. binding constraints missing in the reduced formulation) and false positive (i.e. non-binding constraints predicted as binding) predictions on the meta-loss.
We demonstrate our findings on grids 162-ieee-dtc and 300-ieee with both DC and AC formulations using their default grid parameters. 
First, we solved the full OPF models and determined the binding constraints.
To investigate the effect of false negative predictions, we randomly removed one, two, three, etc., binding inequality constraints from the active set and computed the meta-loss.
For false positive predictions, we extended the active set by a given number of randomly selected constraints from the non-binding set.

\begin{figure}[H]
    \centerline{\includegraphics[width=1.0\textwidth]{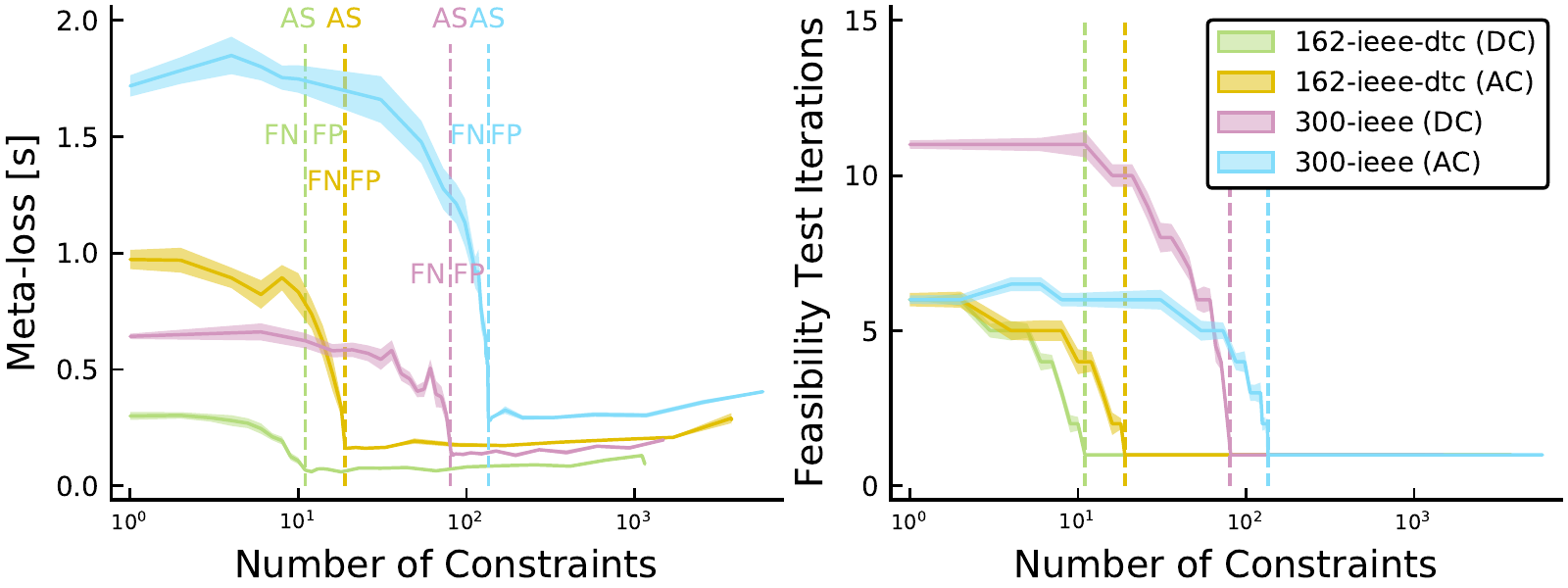}}
    \caption{Profile of the meta-loss (left) and number of iterations within the iterative feasibility test (right) as functions of the number of constraints for two grids, and a comparison of DC vs. AC formulations. 
    Perfect classifiers with the active set (AS) are indicated by vertical dashed lines, false positive (FP) region is to the right and false negative (FN) region is to the left.}
    \label{fig:constraints_vs_cost}
\end{figure}

For each case, we ran $20$ independent experiments and the results are presented in Figure~\ref{fig:constraints_vs_cost}.
The left panel shows the actual meta-loss values, while the right panel presents the number of required iterations in the iterative feasibility test.
For all cases, vertical dashed lines indicate the position of the perfect classification, i.e. the exact active set when no false positive or false negative predictions are present.
When all active constraints are found, including false positive constraints (moving right from the perfect classification) has a marginal effect, however, they slowly but surely increase the computational cost.
The iterative feasibility test converges always within a single step and the cost of the OPF problem depends only on its size.
False negative predictions (moving left from the perfect classification) have much more severe effect: they require more iterations in the feasibility test that significantly increases the meta-loss even in the lack of few active constraints.
Since for small grids the computational cost of the perfect prediction is only $\sim 50\%$ of the full problem (Table~\ref{tab:mag_acdc}), even a few iterations can have a meta-loss exceeding that of the full OPF problem.
In all cases, different constraints represented (or removed) can have a different impact on the meta-loss, particularly in the false negative region where the deviation is larger. 

\subsection{Loss and Meta-loss During Conventional Optimization}
\label{sec:monitoring}
\begin{figure}[H]
    \centerline{\includegraphics[width=1.0\textwidth]{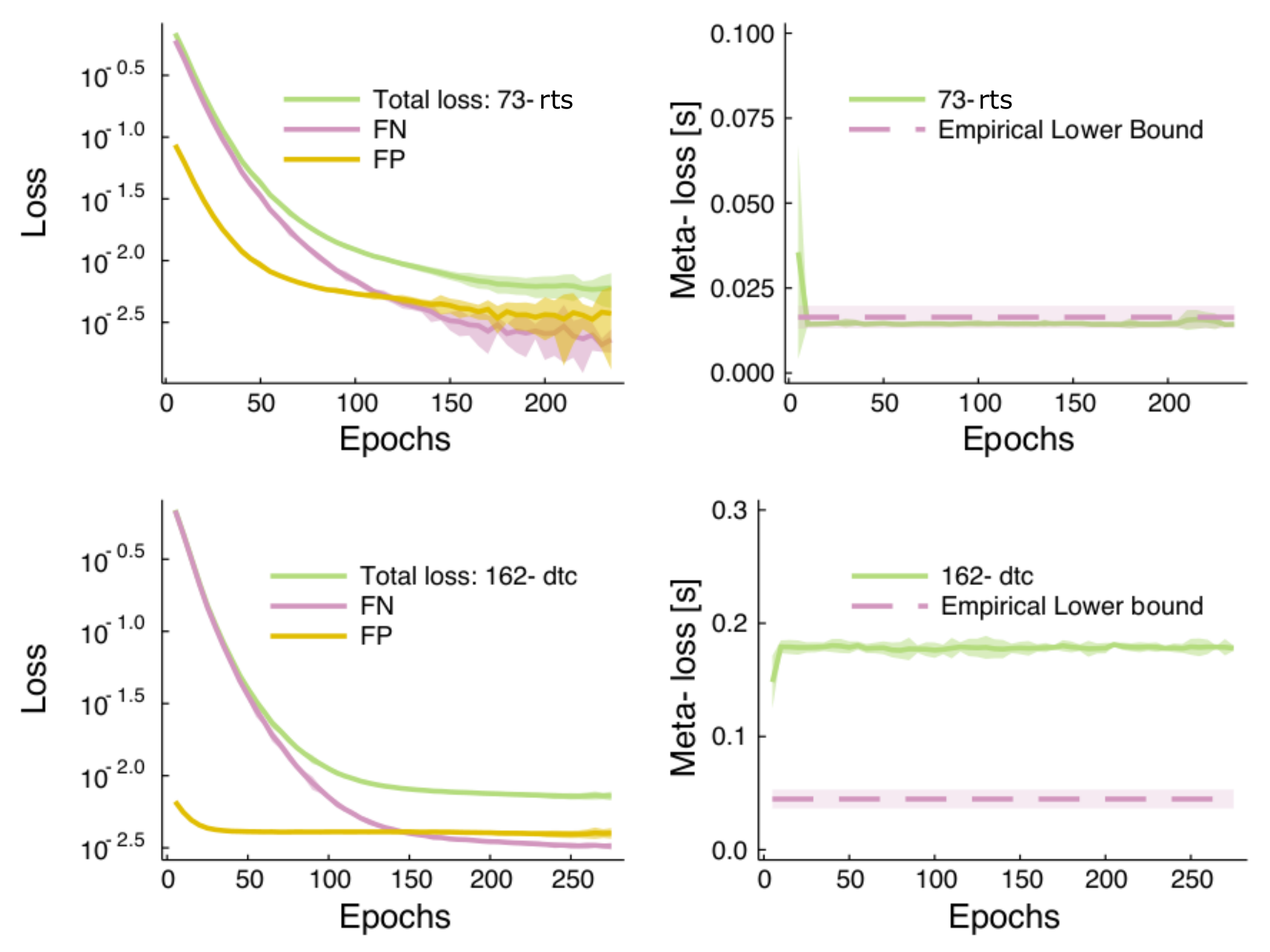}}
    \caption{Loss and meta-loss as functions of epochs during the optimization of the binary cross-entropy objective for a smaller (73-ieee-rts) and a larger (162-ieee-dtc) grid. LHS: each cross-entropy loss is broken down into the constituent false positive (FP) and false negative (FN) contributions. RHS: meta-loss (computational time) evaluated at interval epochs. The empirical lower bound is also plotted representing the typical lower bound for the given grid with perfect reduction (see Table \ref{tab:mag_acdc}).}
    \label{fig:loss_vs_metaloss}
\end{figure}

To demonstrate that conventional loss optimization is not necessarily able to improve the meta-loss, we performed the following experiment on a smaller (73-ieee-rts), and a larger (162-ieee-dtc) grid with DC-OPF formulations using the standard grid parameters on $1$k samples.
During the optimization of cross-entropy, we saved the actual NN weights every $5$ epochs and computed both the loss and meta-loss values on the test set.
The results of 5 independent experiments are collected in Figure~\ref{fig:loss_vs_metaloss}.
For the smaller grid (73-ieee-rts), which has only $8$ distinct active sets in the training data (see Table~\ref{tab:active_sets_dc}), the meta-loss also decreases progressively due to a near-perfect performance of the classifier for such a simple system.
However, for the larger grid (162-ieee-dtc) the meta-loss seems to be insensitive to the optimization of the conventional loss, and stays at a value far from the empirical lower bound.

\subsection{Improving the Meta-loss using Meta-optimization
\label{sec:improving_metaloss}}
Finally, we present our results of the meta-optimization using $10$k and $1$k samples for the DC- and AC-OPF formulations, respectively.
We first carried out a conventional optimization of the cross-entropy loss and starting from this parameterization of the NN we further optimized the meta-loss through PSO.
We computed the accumulated meta-loss of a test set before (\textit{pre}) and after (\textit{post}) the meta-optimization and computed the gain in the meta-loss relative to the full OPF problem in each case. 

First, we review the results for the DC-OPF formulation.
For smaller grids up to grid 73-ieee-rts, we found marginal improvement using meta-optimization.
The reason is similar to what we found in Section~\ref{sec:monitoring}: for such small systems with a limited number of distinct active sets (Table~\ref{tab:active_sets_dc}), the classifiers were able to predict binding constraints almost perfectly and the meta-loss was already close to optimal.

\begin{figure}[H]
    \centerline{\includegraphics[width=1.0\textwidth]{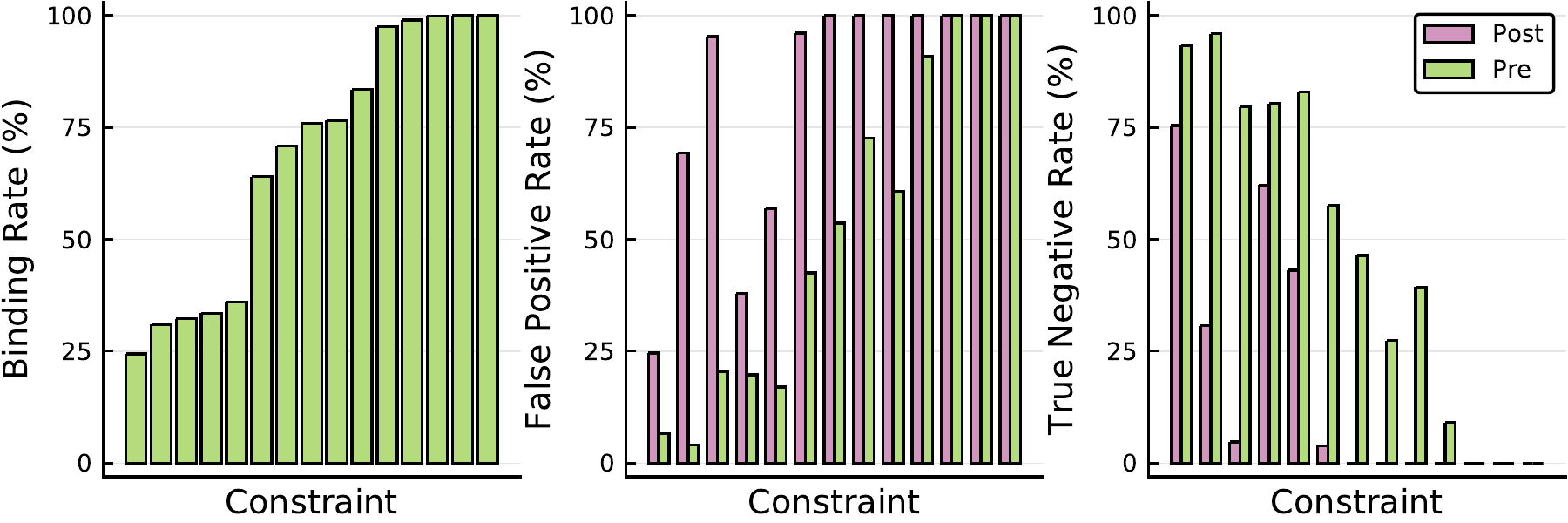}}
    \caption{Distribution of errors pre and post meta-optimization. Left panel: ground truth binding rate on the test set from grid 162-ieee-dtc (using 10k samples). Middle and right panels: comparison of the false positive and true negative rates respectively for pre and post meta-optimization. Constraints are filtered to those that appear at least 20\% in the ground truth. Constraint ordering is the same in each subplot. }
    \label{fig:heatmap_classification}
\end{figure}

For larger systems (from 118-ieee up to 1354-pegase), meta-optimization significantly improved the meta-loss.
However, in many cases, we observed two trivial local minima the meta-optimization could converge to.
The first trivial (Type 1) minimum mostly occurred with smaller training data and the classifier predicted most of the inequality constraints binding.
This is a consequence of the fact that adding false positive predictions to the genuine active set only marginally increases the computational cost as it requires a single feasibility test iteration (Figure~\ref{fig:constraints_vs_cost}).
This results in little signal (via the meta-loss) driving the optimization away from prediction of all constraints binding to the active set.
The second trivial (Type 2) minimum was observed with larger training data, and in this case, the classifier essentially memorized all potentially active constraints in the training set. 
Both are the results of a classifier that has little discriminative power as in each case there is little sensitivity to the actual grid parameters with the optimization learning to allow only a single iteration of the iterative feasibility test. 
Recalling that the maximum achievable gain of the grids we investigated is around 50\% (Table~\ref{tab:mag_acdc}), 
this means that even a single false negative prediction requires an extra iteration of the feasibility test, increasing the total computational time in comparison to the full problem.
For larger grids, we expect a much higher number of possibly binding constraints (Table~\ref{tab:active_sets_dc}), and more significant difference of the meta-loss between the reduced OPF formulations and full model that reduce the possibility of the appearance of these trivial minima.
To avoid the above pathological behavior, we introduced the penalty term discussed in Section-\ref{sec:method_metaoptimization}.
This strategy resulted in a meta-loss-sensitive classifier (Figure~\ref{fig:heatmap_classification}).

The average gains of the meta-loss with two side $95$\% confidence intervals using $10$ independent runs before and after the meta-optimization are collected in the first two columns of Table~\ref{tab:metaloss}.
Gains are computed on the corresponding test sets relative to the meta-loss of the full OPF models as eq.~\ref{eqn:gain} with $t_\mathrm{f} = \sum_{i=1}^{N_{\mathrm{test}}} t_{\mathrm{f}}^{i}$ and $t_\mathrm{ML} = \sum_{i=1}^{N_{\mathrm{test}}} t_{\mathrm{ML}}^{i}$.

For DC-OPF cases, we carried out experiments using $10$k samples.
For 118-ieee, conventional optimization already results in a gain ($38.2$\%) that was improved only slightly by meta-optimization.
Given the limited number of distinct active sets ($122$) this training data size seems to be sufficient to obtain a fairly good classifier using the conventional loss. 
However, as the grid size increases, the gain provided by conventional training becomes drastically worse, resulting in poorer performance compared to that of the full problem.
For each case, meta-optimization was able to improve the meta-loss significantly and bring the gain into the positive regime.

For AC-OPF formulation, 3 grids were investigated using $1$k samples.
As the system size increases, the number of distinct active sets is much larger than in the DC cases, therefore,  it is not surprising that conventional training resulted in a poor gain for all cases.
Meta-optimization again was able to improve all of them into the positive regime.

Finally, we also investigated the effect of the penalty term on the performance of the meta-optimization for case 162-ieee-dtc with the AC-OPF formulation.
Reducing the penalty threshold to the average number of the active constraints in the training data (instead of the default twofold value) significantly increased the false negative predictions resulting in a drop of the gain from $8.6 \pm 7.6\%$ to $-49.7 \pm 4.7\%$.
Using a threshold value of twice of the default one, however, did not improve further the gain ($11.0 \pm 5.9\%$) within the margin of error indicating that the default threshold is already sufficient.  

\begin{table}[H]
\small
\caption{Average gain of classification-based hybrid models in combination with meta-optimization using conventional and weighted binary cross-entropy for pre-training the NN.}
\label{tab:metaloss}
\centering
    \resizebox{\columnwidth}{!}{
    \def\na{\multicolumn{1}{r}{---}}%
    \begin{tabular}{lrcrrcr}
    \toprule
        \multirow{3}{*}{Case} & \multicolumn{6}{c}{Gain (\%)} \\
    \cmidrule(r){2-7}
    & \multicolumn{3}{c}{Conventional} & \multicolumn{3}{c}{Weighted} \\
    \cmidrule(r){2-7}
    & \multicolumn{1}{c}{Cross-entropy} & $\to$ & \multicolumn{1}{c}{Meta-loss} & \multicolumn{1}{c}{Cross-entropy} & $\to$ & \multicolumn{1}{c}{Meta-loss} \\
     \midrule
     DC-OPF ($10$k) \\
     {118-ieee} & $38.2 \pm 0.8$ && $42.1 \pm 2.7$ & $43.0 \pm 0.5$ && $44.8 \pm 1.2$ \\
     {162-ieee-dtc} & $8.9 \pm 0.9$ && $31.2 \pm 1.3$ & $21.2 \pm 0.7$ && $36.9 \pm 1.0$\\
     {300-ieee} & $-47.1 \pm 0.5$ && $11.8 \pm 5.2$ & $-10.2 \pm 0.8$ && $23.2 \pm 1.8$ \\
     {588-sdet} & $-56.0 \pm 0.5$ && $11.9 \pm 9.2$ & $-11.8 \pm 1.0 $ && $24.6 \pm 2.0$ \\
     {1354-pegase} & $-94.6 \pm 2.8$ && $-27.8 \pm 4.7$ & $-54.9 \pm 2.4$ && $-9.9 \pm 5.4$ \\
     \midrule
     AC-OPF ($1$k) \\
     {118-ieee} & $-31.7 \pm 1.2$ && $20.5 \pm 4.2$ & $-3.8 \pm 2.3$ && $29.3 \pm 2.0$ \\
     {162-ieee-dtc} & $-60.5 \pm 2.7$ && $8.6 \pm 7.6$ & $-28.4 \pm 3.0$ && $23.4 \pm 2.2$ \\
     {300-ieee} & $-56.0 \pm 5.8$ && $5.0 \pm 6.4$ & $-30.9 \pm 2.2$ && $15.8 \pm 2.3$ \\    
     \bottomrule
    \end{tabular}
    }
\end{table}

\subsection{Improving the Initial State of Meta-optimization}
Given the importance of a good initialization for meta-optimization we investigated whether further improvement can be attained if the NN weights are initialized at a point closer to a local minimum of the meta-loss. 
Moreover, treating the conventional objective as a surrogate objective~\cite{Goodfellow16} for the meta-loss, we investigated if it can be modified to better represent this. 
For example, we can use a weighted cross-entropy loss that introduces an asymmetry between the false negative and false positive penalty terms:
\begin{equation}
    -y\log\left(\hat{y} \right)w - (1 - y)\log\left(1-\hat{y} \right)(1-w),
    \label{eqn:weighted_binary_crossentropy}
\end{equation}
where $\hat{y} \in [0, 1] $ is the predicted probability of an arbitrary constraint's binding status, $y \in \{0,1\} $ is the ground truth, and $w \in [0, 1]$ is the weight (note that a value of $0.5$ corresponds to the unweighted classical cross-entropy).
As we observed earlier, the meta-loss is much more sensitive to false negative predictions (Figure~\ref{fig:constraints_vs_cost}).
To reflect this in the weighted cross-entropy expression, we carried out a set of experiments for both DC and AC formulations with varying weight (using the same setup for meta-optimization) and found $w=0.75$ as an optimal value for the increased performance.
The results are collected in the second two columns of Table~\ref{tab:metaloss}.
With this modification the gain of pre-training was already improved compared to the conventional cross-entropy, and the corresponding meta-optimization also resulted in further improvement, significantly outperforming the previous results.
For instance, for case 118-ieee, the improved gains came close to the corresponding empirical limits of $47.6 \pm 0.4$ and $31.1 \pm 0.4$ for DC and AC formulations, respectively (Table~\ref{tab:mag_acdc}).

\begin{figure}[H]
    \centerline{\includegraphics[width=1.2\textwidth]{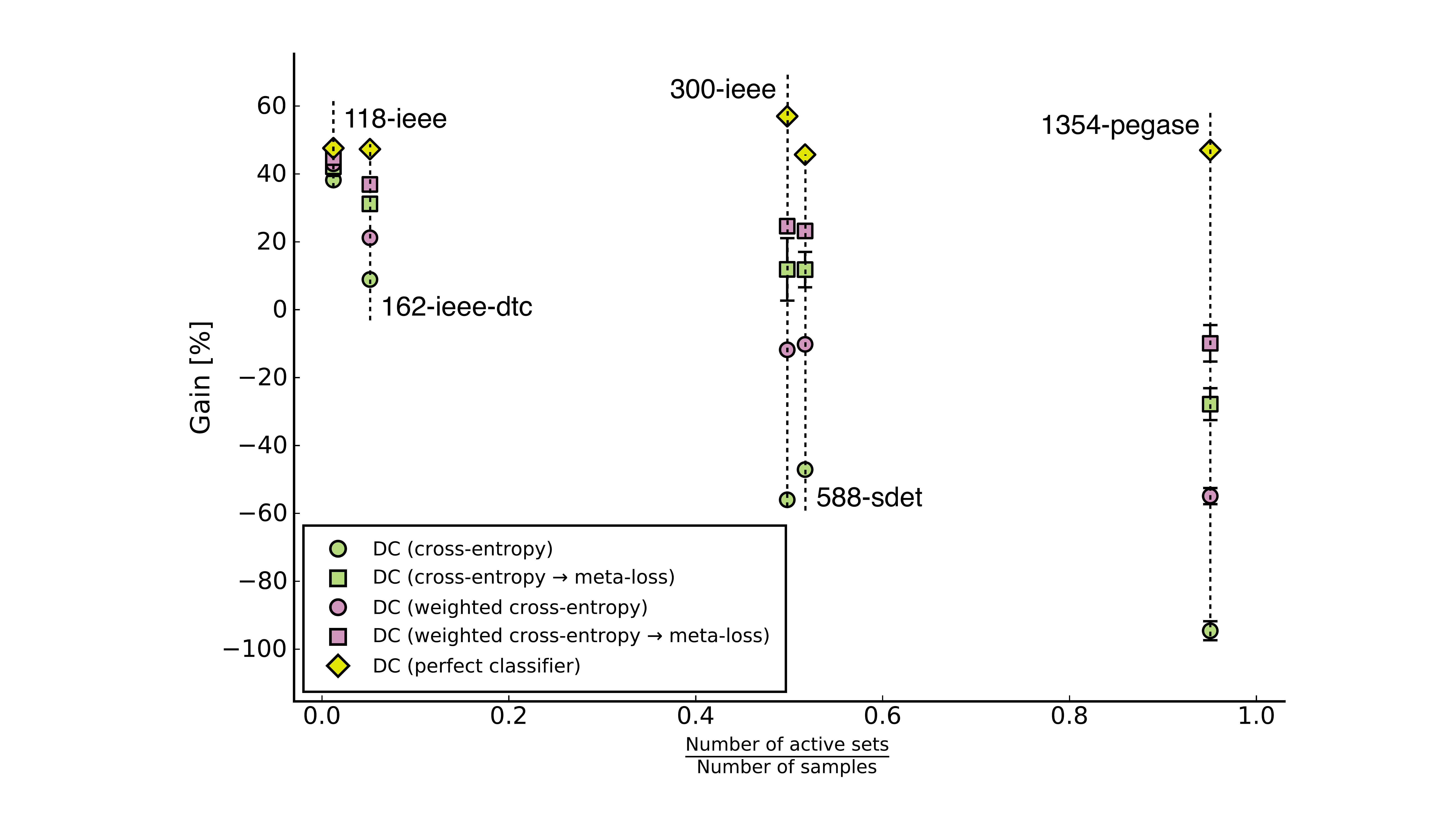}}
    \caption{Correlation between gains of DC formulations obtained by pre-training and subsequent meta-optimization (using conventional or weighted binary cross-entropy loss functions for pre-training) and the ratio of number of active sets and number of training samples. The empirical upper limit of the corresponding gains (i.e. perfect classifier, Table\ref{tab:mag_acdc}) is also shown.}
    \label{fig:gain_vs_active}
\end{figure}

In Figure~\ref{fig:gain_vs_active} we compare the gains of DC formulations obtained before (circles) and after (squares) meta-optimization (using conventional or weighted binary cross-entropy losses for pre-training) as functions of the number of active sets in the training data.
From the figure, it seems that reducing the coverage (i.e. the ratio between the number of active sets and number of samples approaches $1.0$) decreases the gain by meta-optimization.
Therefore, the complexity of the system --- from learning and predictability point of view --- is primarily determined by the potential number of distinct active sets rather than the system size.
Although in general the system size and the number of distinct active sets correlate with each other, there are exceptions: for instance, based on Table~\ref{tab:active_sets_dc}, the size of case 300-ieee is approximately double of the size of 162-ieee-dtc and the size of case 588-sdet is double of 300-ieee.
However, the number of active sets of case 300-ieee is an order of magnitude larger than that of case 162-ieee-dtc, while case 588-sdet has approximately the same.
Accordingly, the achieved gain is very similar for cases 300-ieee and 588-sdet, while case 162-ieee-dtc has a much higher gain.

Finally, we note that an even more representative loss function can be constructed by using individual weights for each constraint.
These weights can then be optimized as hyperparameters using the meta-loss as the optimization target through a similar PSO framework. 
However, our preliminary experiments for DC-OPF showed that although there is a further reduction of the meta-loss, it still required a subsequent meta-optimization of the NN to have competitive performance to the above results. 
This suggests that under this parameterization of the classical objective, although the meta-loss can be minimized to a limited extent, in order to achieve further improvement, a direct meta-optimization of the NN is needed. 
We leave a more thorough exploration to future work.

\section{Conclusion}
A promising approach to reduce the computational time of solving OPF problems is to solve a reduced formulation, which is a considerably smaller problem. 
As part of a classification-based hybrid model, by training models offline, predictions of the active constraint set based on the real-time grid parameters can be performed with negligible cost. 
However, possible false negative predictions and the potential subsequent violation of the corresponding constraints can lead to infeasible points of the original (full) problem. 
This can easily appear for large grids, which have a significant number of distinct active sets. 

This issue can be resolved by the iterative feasibility test used by certain grid operators.
In this procedure, the solution of a reduced OPF problem is tested against all constraints of the full problem, the active set is extended by constraints that are violated, and a new reduced OPF problem is constructed and solved.
The iteration is then terminated when no new constraint is violated, guaranteeing a solution of the full OPF problem.

In this paper, we introduced a hybrid method for predicting active sets of constraints of OPF problems using neural network based classifiers and meta-optimization.
The key ingredient of our approach is to replace the conventional loss function with an objective that measures the computational cost of the iterative feasibility test.
This meta-loss function is then optimized by varying the weights of the NN.

For various synthetic grids, using DC- and AC-OPF formulations we demonstrated that NN classifiers optimized by meta-optimization results in a significantly shorter solve time of the iterative feasibility test than those of conventional loss optimization.
Further, for several DC-OPF cases, the meta-loss as optimized by meta-optimization outperformed that of the full OPF problem.
For AC-OPF, the performance was more moderate due to the large number of non-convex equality constraints, which are responsible for the majority of the computational cost of the OPF calculation.
When comparing the performance for different grid sizes, meta-optimization appears to be an increasingly important component in identifying reduced formulations of OPF problems for larger grids.

We also found that the cross-entropy objective can be modified to obtain an improved meta-loss after conventional training, by weighting the contribution of the two types of classification errors. 
However, particularly for larger-grids, this meta-loss is still higher than that obtained after meta-optimization of the NN parameters directly, indicating that the conventional classification objective is insufficient to capture the meta-loss.  
Nevertheless, these approaches can be straightforwardly used as initialization step for meta-optimization.

Beside increasing the size of the training data set to improve coverage, there are additional ways to improve the method further.
The simplest approach would be to use a more extensive meta-optimization that can be carried out by increasing the number of meta-optimization steps, the number of particles of the PSO method and the number of training sub-samples.
In our experiments, all these parameters had a relatively small value (i.e., 10 particles, 50 iterations and 100 sub-sampled training point) as our intention was to demonstrate that even with this limited setup reasonable results can be achieved.
Using more sophisticated conventional loss functions for pre-optimization is also a promising direction.
In the current approach, all inequality constraints' binding status is predicted, including those that are always non-binding in the training data.
Filtering out these constraints would potentially reduce the output dimensionality of the NN, which could improve the predictive power further.
Finally, we also plan to use topological information of the grid by applying graph neural networks for further improvements.
These changes can make the method suitable for realistic grids and also for $N-k$ contingency problems, where we expect even higher gain due to the smaller fraction of active vs. total number of constraints.

\bibliographystyle{elsarticle-num.bst}
\bibliography{refs.bib}

\begin{thebibliography}{10}
\expandafter\ifx\csname url\endcsname\relax
  \def\url#1{\texttt{#1}}\fi
\expandafter\ifx\csname urlprefix\endcsname\relax\def\urlprefix{URL }\fi
\expandafter\ifx\csname href\endcsname\relax
  \def\href#1#2{#2} \def\path#1{#1}\fi

\bibitem{Tong04}
J.~Tong, Overview of {PJM} energy market design, operation and experience, in:
  2004 IEEE International Conference on Electric Utility Deregulation,
  Restructuring and Power Technologies. Proceedings, Vol.~1, IEEE, 2004, pp.
  24--27.

\bibitem{cain2012history}
M.~B. Cain, R.~P. O’neill, A.~Castillo, History of optimal power flow and
  formulations, Federal Energy Regulatory Commission 1 (2012) 1--36.

\bibitem{datanerc}
NERC, {NERC} reliability guideline, draft, september (2018).

\bibitem{Fiacco68}
A.~V. Fiacco, G.~P. McCormick, Nonlinear Programming: Sequential Unconstrained
  Minimization Techniques, John Wiley \& Sons, New York, NY, USA, 1968,
  reprinted by SIAM Publications in 1990.

\bibitem{Wachter06}
A.~W\"{a}chter, L.~T. Biegler, On the implementation of a primal-dual interior
  point filter line search algorithm for large-scale nonlinear programming,
  Mathematical Programming 106~(1) (2006) 25--57.

\bibitem{Tang17}
Y.~{Tang}, K.~{Dvijotham}, S.~{Low}, Real-time optimal power flow, IEEE
  Transactions on Smart Grid 8~(6) (2017) 2963--2973.
\newblock \href {http://dx.doi.org/10.1109/TSG.2017.2704922}
  {\path{doi:10.1109/TSG.2017.2704922}}.

\bibitem{Liu18}
J.~{Liu}, J.~{Marecek}, A.~{Simonetta}, M.~{Takač}, A coordinate-descent
  algorithm for tracking solutions in time-varying optimal power flows (June
  2018).
\newblock \href {http://dx.doi.org/10.23919/PSCC.2018.8442544}
  {\path{doi:10.23919/PSCC.2018.8442544}}.

\bibitem{zhang2019real}
L.~Zhang, G.~Wang, G.~B. Giannakis, Real-time power system state estimation and
  forecasting via deep unrolled neural networks, IEEE Transactions on Signal
  Processing 67~(15) (2019) 4069--4077.

\bibitem{yang2019two}
Q.~Yang, G.~Wang, A.~Sadeghi, G.~B. Giannakis, J.~Sun, Two-timescale voltage
  control in distribution grids using deep reinforcement learning, IEEE
  Transactions on Smart Grid.

\bibitem{Xavier19}
A.~S. Xavier, F.~Qiu, S.~Ahmed, Learning to solve large-scale
  security-constrained unit commitment problems (2019).
\newblock \href {http://arxiv.org/abs/1902.01697} {\path{arXiv:1902.01697}}.

\bibitem{Guha19}
N.~Guha, Z.~Wang, A.~Majumdar, Machine learning for {AC} optimal power flow,
  ICML, Climate Change: How Can AI Help? Workshop.

\bibitem{Fioretto09}
F.~{Fioretto}, T.~W.~K. {Mak}, P.~{Van Hentenryck}, {Predicting {AC} Optimal
  Power Flows: Combining Deep Learning and Lagrangian Dual Methods}, arXiv
  e-prints (2019) arXiv:1909.10461\href {http://arxiv.org/abs/1909.10461}
  {\path{arXiv:1909.10461}}.

\bibitem{Baker19}
K.~{Baker}, Learning warm-start points for ac optimal power flow, in: 2019 IEEE
  29th International Workshop on Machine Learning for Signal Processing (MLSP),
  2019, pp. 1--6.
\newblock \href {http://dx.doi.org/10.1109/MLSP.2019.8918690}
  {\path{doi:10.1109/MLSP.2019.8918690}}.

\bibitem{Jamei19}
M.~Jamei, L.~Mones, A.~Robson, L.~White, J.~Requeima, C.~Ududec,
  Meta-optimization of optimal power flow, ICML, Climate Change: How Can AI
  Help? Workshop.

\bibitem{Pan19}
X.~Pan, T.~Zhao, M.~Chen,
  \href{http://dx.doi.org/10.1109/SmartGridComm.2019.8909795}{Deepopf: Deep
  neural network for dc optimal power flow}, 2019 IEEE International Conference
  on Communications, Control, and Computing Technologies for Smart Grids
  (SmartGridComm)\href {http://dx.doi.org/10.1109/smartgridcomm.2019.8909795}
  {\path{doi:10.1109/smartgridcomm.2019.8909795}}.
\newline\urlprefix\url{http://dx.doi.org/10.1109/SmartGridComm.2019.8909795}

\bibitem{Zamzam19}
A.~Zamzam, K.~Baker, Learning optimal solutions for extremely fast ac optimal
  power flow (2019).
\newblock \href {http://arxiv.org/abs/1910.01213} {\path{arXiv:1910.01213}}.

\bibitem{Zhou11}
Q.~{Zhou}, L.~{Tesfatsion}, C.~{Liu}, Short-term congestion forecasting in
  wholesale power markets, IEEE Transactions on Power Systems 26~(4) (2011)
  2185--2196.
\newblock \href {http://dx.doi.org/10.1109/TPWRS.2011.2123118}
  {\path{doi:10.1109/TPWRS.2011.2123118}}.

\bibitem{Misra18}
S.~{Misra}, L.~{Roald}, Y.~{Ng}, {Learning for Constrained Optimization:
  Identifying Optimal Active Constraint Sets}, arXiv e-prints\href
  {http://arxiv.org/abs/1802.09639} {\path{arXiv:1802.09639}}.

\bibitem{Pineda20}
S.~{Pineda}, J.~M. {Morales}, A.~{Jiménez-Cordero}, Data-driven screening of
  network constraints for unit commitment, IEEE Transactions on Power Systems
  35~(5) (2020) 3695--3705.
\newblock \href {http://dx.doi.org/10.1109/TPWRS.2020.2980212}
  {\path{doi:10.1109/TPWRS.2020.2980212}}.

\bibitem{ma2009security}
X.~Ma, H.~Song, M.~Hong, J.~Wan, Y.~Chen, E.~Zak, The security-constrained
  commitment and dispatch for {M}idwest {ISO} day-ahead co-optimized energy and
  ancillary service market, in: 2009 IEEE Power \& Energy Society General
  Meeting, IEEE, 2009, pp. 1--8.

\bibitem{Deka19}
D.~{Deka}, S.~{Misra}, {Learning for DC-OPF: Classifying active sets using
  neural nets}, arXiv e-prints (2019) arXiv:1902.05607\href
  {http://arxiv.org/abs/1902.05607} {\path{arXiv:1902.05607}}.

\bibitem{Ng18}
Y.~{Ng}, S.~{Misra}, L.~A. {Roald}, S.~{Backhaus}, {Statistical Learning For DC
  Optimal Power Flow}, arXiv e-prints (2018) arXiv:1801.07809\href
  {http://arxiv.org/abs/1801.07809} {\path{arXiv:1801.07809}}.

\bibitem{Coffrin18}
C.~Coffrin, R.~Bent, K.~Sundar, Y.~Ng, M.~Lubin, {PowerModels.jl}: An
  open-source framework for exploring power flow formulations, in: 2018 Power
  Systems Computation Conference (PSCC), IEEE, 2018, pp. 1--8.

\bibitem{Bishop06}
C.~M. Bishop, Pattern Recognition and Machine Learning (Information Science and
  Statistics), Springer-Verlag, Berlin, Heidelberg, 2006.

\bibitem{Mones18}
L.~Mones, C.~Ortner, G.~Cs{\'a}nyi, Preconditioners for the geometry
  optimisation and saddle point search of molecular systems, Scientific Reports
  8~(1) (2018) 13991.
\newblock \href {http://dx.doi.org/10.1038/s41598-018-32105-x}
  {\path{doi:10.1038/s41598-018-32105-x}}.

\bibitem{Kennedy95}
J.~{Kennedy}, R.~{Eberhart}, Particle swarm optimization, in: Proceedings of
  ICNN'95 - International Conference on Neural Networks, Vol.~4, 1995, pp.
  1942--1948 vol.4.
\newblock \href {http://dx.doi.org/10.1109/ICNN.1995.488968}
  {\path{doi:10.1109/ICNN.1995.488968}}.

\bibitem{Zhan09}
Z.~{Zhan}, J.~{Zhang}, Y.~{Li}, H.~S. {Chung}, Adaptive particle swarm
  optimization, IEEE Transactions on Systems, Man, and Cybernetics, Part B
  (Cybernetics) 39~(6) (2009) 1362--1381.
\newblock \href {http://dx.doi.org/10.1109/TSMCB.2009.2015956}
  {\path{doi:10.1109/TSMCB.2009.2015956}}.

\bibitem{modpso}
L.~Mones, Modified {PSO} in {Optim.jl}, \url{https://github.com/molet/Optim.jl}
  (2019).

\bibitem{Baba19}
S.~{Babaeinejadsarookolaee}, A.~{Birchfield}, R.~D. {Christie}, C.~{Coffrin},
  C.~{DeMarco}, R.~{Diao}, M.~{Ferris}, S.~{Fliscounakis}, S.~{Greene},
  R.~{Huang}, C.~{Josz}, R.~{Korab}, B.~{Lesieutre}, J.~{Maeght}, D.~K.
  {Molzahn}, T.~J. {Overbye}, P.~{Panciatici}, B.~{Park}, J.~{Snodgrass},
  R.~{Zimmerman}, {The Power Grid Library for Benchmarking AC Optimal Power
  Flow Algorithms}, arXiv e-prints (2019) arXiv:1908.02788\href
  {http://arxiv.org/abs/1908.02788} {\path{arXiv:1908.02788}}.

\bibitem{bezanson2012julia}
J.~Bezanson, S.~Karpinski, V.~B. Shah, A.~Edelman, Julia: A fast dynamic
  language for technical computing, arXiv preprint arXiv:1209.5145.

\bibitem{innes2018flux}
M.~Innes, Flux: Elegant machine learning with {J}ulia., J. Open Source Software
  3~(25) (2018) 602.

\bibitem{Ioffe15}
S.~Ioffe, C.~Szegedy, Batch normalization: Accelerating deep network training
  by reducing internal covariate shift, in: F.~Bach, D.~Blei (Eds.),
  Proceedings of the 32nd International Conference on Machine Learning, Vol.~37
  of Proceedings of Machine Learning Research, PMLR, 2015, pp. 448--456.

\bibitem{Vinod2010ICML}
V.~Nair, G.~E. Hinton, Rectified linear units improve restricted boltzmann
  machines, in: Proceedings of the 27th International Conference on
  International Conference on Machine Learning, ICML’10, Omnipress, Madison,
  WI, USA, 2010, p. 807–814.

\bibitem{Srivastava2014JMLR}
N.~Srivastava, G.~Hinton, A.~Krizhevsky, I.~Sutskever, R.~Salakhutdinov,
  Dropout: A simple way to prevent neural networks from overfitting, J. Mach.
  Learn. Res. 15~(1) (2014) 1929–1958.

\bibitem{kingma2014adam}
D.~P. Kingma, J.~Ba, Adam: A method for stochastic optimization, arXiv preprint
  arXiv:1412.6980.

\bibitem{mogensen2018optim}
P.~K. Mogensen, A.~N. Riseth, Optim: A mathematical optimization package for
  {J}ulia, Journal of Open Source Software 3~(24) (2018) 615.
\newblock \href {http://dx.doi.org/10.21105/joss.00615}
  {\path{doi:10.21105/joss.00615}}.

\bibitem{domahidi2013ecos}
A.~Domahidi, E.~Chu, S.~Boyd, {ECOS}: An {SOCP} solver for embedded systems,
  in: 2013 European Control Conference (ECC), IEEE, 2013, pp. 3071--3076.

\bibitem{osqp}
B.~Stellato, G.~Banjac, P.~Goulart, A.~Bemporad, S.~Boyd, {OSQP}: An operator
  splitting solver for quadratic programs, ArXiv e-prints\href
  {http://arxiv.org/abs/1711.08013} {\path{arXiv:1711.08013}}.

\bibitem{Goodfellow16}
I.~Goodfellow, Y.~Bengio, A.~Courville, Deep Learning, MIT Press, 2016.

\end{thebibliography}

\end{document}